\documentstyle[sprocl]{article}

\def\be{\begin{equation}}
\def\ee{\end{equation}}
\def\bea{\begin{eqnarray}}
\def\eea{\end{eqnarray}}
\begin{document}

\title{Why the Quark-Gluon Plasma Isn't a Plasma}

\author{Robert D. Pisarski}

\address{Department of Physics,\\Brookhaven National Laboratory\\
Upton, NY 11973 USA\\e-mail: pisarski@bnl.gov}


\maketitle
\abstracts{An alternate picture of the deconfined phase of gauge
theories is described.  Instead of a plasma, the theory is
viewed as a condensate of Polyakov
lines.  The pressure is determined by an elementary mean field theory.}

\begin{center}
{\it To appear in the Proceedings of: Strong and Electro-Weak Matter 2000}\\
{\it C. P. Korthals-Altes, editor}
\end{center}
\section{Introduction}

A major reason for studying the collisions of
heavy nuclei at very high energies is
the hope that we might observe a new state of matter, the 
``Quark-Gluon Plasma''.  In this Proceeding I outline a different
picture for the high temperature phase of $SU(N)$ gauge theories: it isn't a
plasma --- which in this case means 
a weakly interacting gas of essentially massive
quasi-particles --- but a condensate of $Z(N)$ spins.  

That the high temperature phase of a $SU(N)$ gauge theory is like the low
temperature phase of a $Z(N)$ spin model is
an old and familiar story.  The role of a global $Z(N)$ 
symmetry, and its relationship to confinement, was first introduced
by 't Hooft.~\cite{hooft}  That $Z(N)$ spins form a rigorous 
order parameter for the deconfining phase transition in pure
glue theories was shown by Polyakov and by Susskind.~\cite{polyss}  
Svetitsky and Yaffe~\cite{svet} demonstrated that
if the deconfining phase transition is of second order, that 
the $Z(N)$ spins determine the universality class of the theory.
Indeed, the $Z(N)$ spins are
the standard order parameter used on the Lattice to pinpoint the
deconfining phase transition.  

What is novel in the following discussion is a mean field theory
which relates the pressure to the 
behavior of the $Z(N)$ spins.  I first
review the usual quasi-particle picture of the high temperature phase,
and then discuss how I was led to the model of 
a $Z(N)$ condensate.  The advantage of a mean field theory
is that while it clearly breaks down at a critical point, experience
in condensed matter systems tells us that it often 
works well away (and sometimes even relatively near) 
a critical point.  While the model is
extremely simple, it does make numerous detailed predictions,
which can be tested by numerical simulations.  My presentation is
meant to be pedagogical, ignoring as many technical details as possible.

\section{Quasi-particle Models}

I largely consider the theory without dynamical quarks.  Of course
this is a serious limitation, and yet I will argue that in fact
the quenched limit can tell us interesting things, at least for
three colors.

At very high temperatures, by asymptotic freedom a plasma
picture is bound to be a good description.
As the temperature $T \rightarrow \infty$, the pressure
$p \sim n_\infty T^4$; $n_\infty$ counts the numbers of
degrees of freedom in a textbook fashion.  There are perturbative 
corrections to this result; the series starts with terms $\sim O(g^2)$,
but due to many-body effects, the next term is $\sim O(g^3)$, so the
correct series is in fact an expansion in $\sqrt{g^2}$.  Heroic
calculations~\cite{pert} have given us the series to $\sim O(g^5)$.

Unfortunately, the series behaves extremely badly.  Balancing just
the $g^2$ and $g^3$ terms tells one that the series starts to fail
when $\alpha = g^2/(4 \pi) \sim 1/20$, which corresponds to
ridiculously high temperatures, $\sim 10^7$GeV or so.  
This series has been studied with 
Pade approximations;\cite{pade} Parwani claims that useful information
can be extracted, which is possible, given the number of terms
computed.

Even so, it is not difficult to understand why naive perturbation
theory may fail.  In ordinary
perturbation theory, one is computing about the trivial, perturbative
vacuum, in which all fields propagate at the speed of light.  Through
scattering in the thermal bath, however, particles develop an index
of refraction different from unity, which heuristically we call
a ``thermal mass''.  (While handy, the nomenclature is misleading, because
gauge invariance is not violated.)  The thermal mass of the gluon
is~\cite{debye}
\begin{equation}
m_g = \frac{\sqrt{N}}{3} \; g T \; .
\label{e1}
\end{equation}
In fact, the series is an expansion in $\sqrt{g^2}$ precisely because
of these thermal masses.  

Thus it is useful to develop an expansion in which the
effects of thermal masses are automatically included.  
Phenomenological approaches to gases of massive quasiparticles
have been developed.\cite{quasi}
More recently, approaches based upon hard thermal loops were
developed by two groups, by Andersen, Braaten, and
Strickland, and by Blaizot, Iancu, and Rebhan.~\cite{htl}  There
are crucial differences between these two approaches: the 
approach of Andersen et al.
is manifestly gauge invariant, but even reproducing
the $\sim g^2$ term in the free energy requires a two (!) loop
calculation in an effective theory.  The approach of the Blaizot
et al gets the $\sim g^2$ term right, and with some sleight
of hand, the $\sim g^3$ term as well, but it is not manifestly gauge
invariant.  After these resummations, the free energy is presumably
better behaved than the ordinary perturbative expansion.  (Although
due to the complexity of the calculations, 
the really crucial test of computing higher order corrections
cannot be carried out.)
Both groups agree that 
$p/T^4$ appears to be constant down to low temperatures, several
times the transition temperature, $T_c$; as discussed below, this
appears to agree with Lattice simulations.

Let us take the formula for the thermal mass, $m_g$, in Eq. (\ref{e1})
seriously.  Coming down from infinite temperatures, the ratio
of $m_g/T$ grows slowly as the temperature decreases, due simply to the
logarithmic increase of the QCD 
coupling constant.  This agrees with 
careful studies on the Lattice, such as by Laine
and Philipsen.~\cite{laineph}

\section{Lattice results}

The interesting question is to match the behavior of the free energy
down to $T_c$.  Here one must appeal to numerical simulations on the
Lattice.~\cite{lattice_three}  I stress that the data in the quenched
limit is extremely close to the continuum limit; for example, between
different groups, the
results for the ratio of the critical temperature, to the square
root of the string tension, agree to within $\sim 5\%$; the difference
is due to how the string tension is extracted, not $T_c$
{\it per se}.  The
Lattice tells us several things.  First, that $p/T^4$ is 
{\it constant} from a temperature $\kappa T_c$ on up, 
equal to about $80\%$ of the ideal gas value; $\kappa \approx 2$.
The pressure drops
suddenly from $\kappa T_c$ to $T_c$, and is essentially {\it zero}
below $T_c$.  This reflects the fact that $T_c \sim 265$~MeV is low
in the pure glue theory; at such temperatures, the probability to
excite glueballs, which are heavy, $\geq 1.5$~GeV, is very small.

The standard way of understanding this behavior of the pressure is
to use a bag model.  The pressure is then $p = n_\infty T^4 - b$,
where $b$ is the bag constant.  By adjusting $b$, one can obviously
get the pressure to vanish at some point.  

This is in direct contradiction to other data given to us by 
the Lattice.  In particular, it is known that
for three colors, the deconfining phase transition is {\it weakly}
first order.  For example, the
ratio of the energy at $T_c$, to the ideal gas value, is $\sim 1/3$.
This is very different from the bag equation of state, which by
construction gives a similar ratio $=1$.  

The latent heat, however, underestimates how much correlation lengths
grow near the deconfining transition temperature, 
$T_c$.~\cite{potentials}  
Coming to the transition from below, 
the string tension at $T_c$ is a factor of {\it ten}
smaller than at zero temperature.  Coming to the transition
from above, the ratio of the screening mass 
(as defined from the two point function of Polyakov loops), divided
by the temperature, decreases by a 
factor of {\it ten}, as one goes from $2 T_c$ to just above $T_c$.

A historical comment is in order.  Originally, simulations for three
colors at a small number of lattice steps in the imaginary time
direction, $n_t$, found a strong first
order deconfining transition.  Going to larger $n_t$, at first the
APE group claimed to find a {\it second} order transition, in direct
contradiction with very general arguments for a 
first order transition.~\cite{svet}  While it was soon found by the
Columbia group that the transition is first order, the confusion
was extremely
instructive, for it demonstrated how {\it weakly} first order it
really is.  

The order of the phase transition for three colors can be understood
as follows.  For two colors, the deconfining transition appears,
from numerical simulations, to be of second order.  Indeed, calculations
of critical exponents agree very well with the prediction~\cite{svet}
of a $Z(2)$, or Ising, spin model.\cite{two}  For four colors,
old simulations indicated that the transition is of first order.
For various reasons, Tytgat and I were skeptical of this result,~\cite{wrong}
but we were wrong.
In agreement with older simulations, recent 
work by Ohta and Wingate clearly demonstrates
that the transition {\it is} of first order.~\cite{ohta} 
(For four colors, large $n_t$ is necessary to seperate a
bulk transition of the Wilson action from the
deconfining transition at nonzero temperature.)
There is still work to do, though: it is not known how strongly
first order the deconfining transition is in the continuum limit.
If we assume that it is strongly first order, then the transition
for three colors is weakly first order because it is near the 
second order transition for two colors.  Presumably,
four colors is like any larger number of colors, so that the
transition is of first order as $N\rightarrow \infty$.  This agrees
with general arguments on the Lattice by Gocksch and Neri.~\cite{gock}
Also, it demonstrates that in at least this one instance, three colors
is {\it not} close to infinity, but to two.  

\section{Back to Quasiparticles}

How can a weakly first order transition be described within
a quasiparticle model?  
Take non-interacting quasiparticles, with masses which vary with
temperature.  Since the particles
are by definition noninteracting, in order for the pressure
to {\it nearly} vanish at $T_c$, the quasiparticle masses must be
{\it heavy}, so that their free energy vanishes in a Boltzmann fashion.
This has been done by several groups.~\cite{quasi}
Typically,
the one loop formula for the thermal mass in Eq. (\ref{e1}) is
assumed; one then allows the coupling constant to grow in such a fashion that
the pressure is correctly fit.  Essentially, one correlates the
Landau pole in the coupling constant with the decrease in the free
energy at $T_c$.  (For a different, phenomenological fit to the pressure, 
see.~\cite{fit})

I suggest that this fit is physically unnatural.  Consider,
for the sake of argument, the case of two colors.  Again,
the pressure, the energy, {\it etc.}, nearly vanish at
$T_c$.  Surely the pressure for two colors can be fit by such
a model, with increasing masses.  We know, however, that since
the transition is truly of second order, that the mass for
$Z(2)$ spins exactly {\it vanishes} at the critical temperature.
So we are describing an exact critical point by a gas in which 
the masses are becoming very heavy, not even light!  I find
this odd; one would like a model to fit all aspects
of the physics, not just some.  The same argument applies for three colors,
where the transition is weakly first order, or as I prefer to
call it, nearly second order.  Conversely, 
if the deconfining transition
is strongly first order for four or more colors, then such a fit may
be reasonable.

I note that the decrease of the screening mass was described
by Peshier {\it et al}.\cite{quasi} 
There are actually two screening masses: one for the screening of
static, electric fields, $\sim A_0^2$, and
one for the screening of time-dependent fields, $\sim A_i^2$.
In ordinary perturbation theory, these two masses are strictly tied
together, since they both arise from the gluon polarization tensor,
with one $1/\sqrt{3}$ times the other.  
Peshier {\it et al}~\cite{quasi} describe the mass
for $A_0$ as an integral over the $A_i$ quasiparticles; as the 
$A_i$ fields become heavy, the $A_0$ fields become light.
It is not clear to
me, though, how these
two very different screening masses can arise from the same
polarization tensor.  See, however, my comments at the end.

\section{Effective Model}

How then can one describe the behavior of the pressure?  The impetus
for my work was a talk given by Eric Braaten at
the Aspen Center for Physics, in August of 1999.  
After describing his work,~\cite{htl} 
Hans Pirner asked him if he had considered
the coupling of condensate fields to the quasiparticles; Braaten
said no.  (I think Pirner meant pions.)
My immediate thought was to couple the field for the Polyakov
loop to the quasiparticle masses.  

I then introduce the $Z(N)$ Polyakov loop:
\begin{equation}
\ell(x) = \frac{1}{N} \; {\rm tr}\left(
{\cal P} \exp \left( i g \int^{1/T}_0 A_0(x,\tau) \, 
d\tau \right) \right)\; ;
\label{e2}
\end{equation}
where $\cal P$ is path ordering, $g$ is the gauge coupling constant,
$x$ is the coordinate for three spatial 
dimensions, and $\tau$ that for euclidean time.  The $\ell$ field
is real for two colors, and complex valued for three or more.
In published work 
I call the operator in (\ref{e2}) the thermal Wilson line;~\cite{rp1}
in these proceedings, in deference to European custom I call it
the Polyakov loop.~\cite{polyss}

The $\ell$ field transforms only under
the global $Z(N)$ symmetry of 't Hooft,~\cite{hooft} $\ell \rightarrow
exp(2\pi/N) \ell$.  The expectation value of $\ell$, $\langle \ell
\rangle = \ell_0$, behaves exactly opposite to that of an ordinary
spin system: it vanishes in the low temperature phase, and is
nonzero in the high temperature phase.  

My first thought was then to couple the Polyakov loop to the screening
mass, taking $m_g\sim g T \ell$ for gluons, and $m_q \sim g T \sqrt{\ell}$
for quarks.  These powers of $\ell$ are natural guesses, given the
general expressions for hard thermal loops.\cite{lebellac}  However,
the quasiparticles then become light near the transition, and even
if one adds a bag constant to fit the pressure, one can't fit the
energy, since light fields inevitably have a lot of energy.

Instead of marrying the Polyakov loop to hard thermal loops, I ended
up throwing out the quasiparticles altogether, and developing a theory
entirely in terms of the Polyakov loop.
The usual way in which to model the magnetization in a spin system
is by mean field theory.  For three colors, I take:
\begin{equation}
{\cal V} \; = \;
\left( - 2 b_2 \, |\ell|^2
+ b_3 ( \ell^3 + (\ell^*)^3 )
+ (|\ell|^2)^2 \right) \; b_4 \; T^4 \; .
\label{e3}
\end{equation}
The terms $\sim |\ell|^2$ and $\sim (|\ell|^2)^2$ appear for
any number of colors, since they are invariant under a global
symmetry of $O(2)$.  The term $\sim \ell^3$ is special to the
three colors, reducing the global symmetry to $Z(3)$.  
The cubic invariant drives the
transition first order; for the sake of simplicity, since
the transition is nearly second order, for now I ignore $b_3$.

At the minimum of the potential, for $b_3 =0$,
$\ell_0^2 = b_2$.  Thus whatever the behavior of $\ell_0(T)$, one can
trivially obtain this by
adjusting $b_2(T)$.  In an ordinary spin system, $b_2$ is negative
below $T_c$, and positive above; for $Z(3)$ spins in a gauge
theory, one simply requires the
opposite, positive $b_2$ below $T_c$, and negative $b_2$ above.

The potential in (\ref{e3}) is totally standard;~\cite{svet} 
my only contribution
is to add an overall factor of $T^4$ in 
$\cal V$.  This arises because the Polyakov loop,
as an exponential, is of necessity a pure number, without any
mass dimension.  Thus the only parameter to make up powers of
mass is the temperature $T$.  While mathemtically trivial, this
does have physical consequences.

At high temperature, fluctuations in $A_0$ can be neglected,
so $\ell_0 \rightarrow 1$.  With $b_3=0$, this requires
that $b_2 \rightarrow 1$ as $T \rightarrow \infty$.  
At the minimum of the potential, when $b_2 < 0$,
${\cal V} = - b_2^2 b_4 T^4$; when $b_2 > 0$, 
${\cal V} = 0$.  Since the pressure $p = - {\cal V}$, we
obtain a relation between the pressure and $\ell_0$ in the
deconfined, or broken symmetric phase, $p \sim \ell_0^4 T^4$.
At high temperature, $b_4 \rightarrow n_\infty$, to obtain
the ideal gas result.
As the temperature is lowered, certainly $b_2$ changes,
and probably $b_4$ as well.  Around the critical temperature,
in the spirit of mean field theory I assume that one can
neglect the variation in $b_4$, and only allow $b_2$ to vary.

The relationship between the pressure and $\ell_0(T)$ can in
principle be tested on the Lattice.  Qualitatively, one
views $p/T^4$ as being approximately constant from $T=\infty$
down to $\kappa T_c$ because $\ell_0$ is approximately constant.
From the mean field theory alone, one cannot tell if 
$p/T^4$ is $80\%$ of the ideal gas value because $b_4$ is 
$80\%$ of $n_\infty$, or because $\ell_0$ is $.95$ (or,
equivalently, because $b_2 \sim .9$).  
Indeed, the best way of defining $\kappa$ is not merely from
the free energy, but where $\ell_0$ differs significantly from
unity.  Below $\kappa T_c$, the sharp drop in the pressure is then viewed
as the result of a sharp drop in the expectation value of the
$Z(3)$ Polyakov loop.   

In principle, this can be tested by measuring $\ell_0$ on the
Lattice.  For example, it is not obvious that the simple quartic
potential in $\cal V$ is necessarily adequate to fit the
pressure; perhaps terms $\sim (|\ell|^2)^3$ have to be added.
If only quartic terms enter, then one has a qualitative prediction:
$p/(\ell_0^4 T^4)$ should be constant, at all temperatures.  
Of course near $T_c$ this may break down because of nearly
critical fluctuations.  This is certainly true for two colors.

Two natural questions arise, which have a related answer.
First, why in the mean field theory is there a contribution from
the condensate, but not from fluctuations in the condensate?
Second, is this type of mean field theory for the pressure 
useful for any gauge theory at nonzero temperature?

To answer the first question, I appeal to the large $N$ 
expansion.~\cite{thorn,wrong}
At large $N$, the free energy itself is a natural order parameter:
it is of order one in the confined phase, and of order $\sim N^2$
in the deconfined phase.~\cite{thorn}  The elementary explanation
for this is of course that gluons are deconfined, and they contribute
$\sim N^2$ to the free energy.  However, there is a quandry:
we should be able to describe the free energy, which is a physical
quantity, exclusively in terms of gauge invariant quantities.
So what is the term in the free energy $\sim N^2$ due to?  
I suggest that the expectation value of the $Z(N)$ Polyakov loop,
$\ell_0$, is the only quantity which can do so.  The potential
is like $\cal V$ in Eq. (\ref{e3}), except that
there is no $\ell^3$ term, only terms $\sim |\ell|^2$ and
$(|\ell|^2)^2$.  With the normalization of the $Z(N)$ Polyakov
line in Eq. (\ref{e1}), 
$\ell_0$ is of order one (not $\sim N$) at all temperatures.  
So one simply adjusts the overall coefficient, $b_4$, to be
$\sim N^2$, as it must be to match the ideal gas value.  
There are then other contributions to the free energy:
from fluctuations in $\ell$, for example.  But $\ell$ only has
two degrees of freedom, and so can be ignored relative to the
terms $\sim N^2$.  There are also contributions to the free energy
from glueballs, either electric or magnetic.  Again, however,
any glueball state is of necessity a color singlet, and so
can only contribute $\sim 1$ to the free energy.  

The $Z(N)$ Polyakov loop does not always dominate the pressure.
Consider a generalized (or ``quarky'') large $N$ limit,
in which the number of flavors, $N_f$, goes to infinity along
with $N$.  Since there are $\sim N_f^2$ hadrons, the free energy
is then nonzero, on a scale of $\sim N_f^2 \sim N_f N \sim N^2$,
at all temperatures.  Alternately, the expectation value of the
$Z(N)$ Polyakov loop, $\ell_0$, is always nonzero: $\sim
N_f/N$ at low temperatures, and $\sim 1$ at high temperatures.
One might still imagine that there is a phase transition (for
massless quarks, there is a chiral phase transition), but one
does not necessarily expect a large change in the free energy,
nor in $\ell_0$.  One could still write down a mean field theory
for the pressure, in terms of a potential like $\cal V$, but it
is not obvious if it will be of any use.

Indeed, this gives one a qualitative guide to when the mean field
theory for $\ell$ should be of use.  {\it If} the pressure in the
low temperature phase is small, then presumably that is because
$\ell_0$ is small.  One can then model the increase in the free energy
by a change in the Polyakov loop.  Fortunately, this appears to be
the case in QCD: from present Lattice simulations,\cite{karsch2}
the pressure in the low temperature phase {\it is} small.  Of course,
in present simulations, the pions are relatively heavy, at least
as heavy as physical kaons.  This is where the present theory could
fail to apply to QCD: if better simulations, with lighter pions,
find that the free energy is much larger in the low
temperature phase than at present, then
using a mean field theory for $\ell$ to calculate the pressure
would become increasingly dubious.

For the time being, let me assume that this is {\it not} the case.  Now
it is well known that the order of the phase transition depends
{\it extremely} sensitively on the presence of dynamical quarks;
it appears that the weakly first order transition in the pure
glue theory is completely washed out by the effect of quarks.
I suggest that this is only possible because the transition is
so weakly first order.  I suggest, however, that the nearly second
order transition persists.  In the pure glue theory,
correlation lengths associated with $\ell$ grow by a factor
of ten near the deconfining phase transition.  With dynamical
quarks, it is no longer a deconfining phase transition, but
a chiral transition, or just crossover.  Thus one does not
expect correlation lengths associated with $\ell$ to grow
as much; even within the model, they cannot, since then
there is measurable pressure in the low temperature phase.
Perhaps the correlation length associated with $\ell$ might
grow not be a factor of ten, but five.  This factor of five
is sheer guesswork.

What is dramatic is that this field may dominate the free energy in QCD.
In an ordinary critical point, there is both a regular and a singular
part.  The critical field affects the singular part, but there is always
a regular part which remains constant about the critical temperature.
In QCD, it might be that the regular part --- associated with fields
other than $\ell$, or with fluctuations in $\ell$ --- are much smaller
to those from the potential for $\ell$.  Then it would be
meaningful to speak of the transition in QCD as being ``near'' the critical
point for two colors in the quenched theory.  
It would be preposterous to suggest this without data from the Lattice;
with such data, it is a testable proposition.

\section{Qualms}

In this section I mention several outstanding problems 
which I swept under the rug.

Like any other operator
in a quantum field theory, the Polyakov loop must be renormalized.
It is necessary to seperate this ultraviolet renormalization
from the physical behavior of $\ell$,
as a $Z(N)$ spin.  For example, when $\ell_0$ is measured on
the Lattice, it typically is {\it much} less than unity, even
far away from $T_c$.~\cite{kajantie}  This is due to a renormalization
on the Lattice, which acts to decrease the bare $\ell_0$.~\cite{rp1}

To seperate the effects of ultraviolet renormalization, 
it is probably is not sufficient to measure just $\ell$.
One possibility is the following.~\cite{rp1}
Calculate not just the trace,
but the full eigenvalue distribution for the matrix values, 
$SU(N)$ Polyakov loop.
By definition, one ignores an overall constant factor, which
includes any ultraviolet renormalization.  
The trace of the Polyakov loop can then be computed from the 
distribution of eigenvalues.  By
construction, this gives an $\ell_0$ which approaches unity
at high temperature.  This may or may not be a practial proceedure;
regardless, since one has a renormalization condition with which to
fix $\ell$ --- such as $\ell_0 \rightarrow 1$ as $T \rightarrow \infty$, 
the problem should be manageable.

Recently, the role of the magnetic $Z(N)$ symmetry has been stressed
by several authors.\cite{magnetic}  This is an interesting approach;
however, the magnetic $Z(N)$ symmetry is broken in the low temperature phase,
and restored in the high temperature phase.  But then a magnetic condensate
cannot be used to develop a mean field theory for the free energy
in a gauge theory, since that is only large at high temperatures.

More generally, several authors, most cogently Smilga,\cite{smilga}
have argued that the Polyakov loop
it is entirely a figment of imaginary time.  (Indeed, as was
clear in the discussion section, this
feeling appeared to be shared by many participants at the conference.)
The problems which Smilga raises are important, and I cannot solve
them at present.  However, I believe that it is
possible to analytically continue the Polyakov loop
to processes in real time.  (I made one, incomplete, attempt.~\cite{rp1})
Indeed, as is obvious from the original 
derivation by Taylor and Wong.\cite{taylor}
hard thermal loops\cite{lebellac} generalize
the Debye mass term, $\sim A_0^2$ to real time.  
But the Debye mass term is the leading term in an 
expansion for an effective lagrangian of Polyakov loops.~\cite{rp1}

More generally, perhaps this can reconcile my approach with that
of massive quasiparticles.~\cite{quasi,htl}  
The Polyakov loop is an operator involving
$A_0$, while the quasiparticles involve the transverse degrees of 
freedom, the $A_i$'s.  Maybe near the transition, where $\ell_0
\rightarrow 0$, the $A_i$ propagator is more involved than just a
simple mass getting heavy: maybe it is that the wave function renormalization
becomes small.  After all, $\ell$ is something like the wave function
for an infinitely heavy test quark.

\end{document}